\documentclass[prl,twocolumn,superscriptaddress,showpacs,floatfix]{revtex4}

\usepackage{dcolumn,amsmath,xspace}

\usepackage{graphicx}
\usepackage{dcolumn}
\usepackage{amsmath}

\usepackage{units}

\newcommand{\degC}{\ensuremath{~^{\circ}\text{C }}}

\begin{document}
\title{First direct observation of a nearly ideal graphene band structure\\}

\author{M. Sprinkle}
\affiliation{Georgia Institute of Technology, Atlanta, Georgia
30332-0430, USA\\}
\author{D. Siegel}
\affiliation{Dept. of Physics, Univ. of California, Berkeley, CA 94720, USA\\}
\affiliation{Advanced Light Source, Lawrence Berkeley National Laboratory, Berkeley, CA 94720, USA\\}
\author{Y. Hu}
\affiliation{Georgia Institute of Technology, Atlanta, Georgia
30332-0430, USA\\}
\author{J. Hicks}
\affiliation{Georgia Institute of Technology, Atlanta, Georgia
30332-0430, USA\\}
\author{P. Soukiassian}
\affiliation{Commissariat a l'Energie Atomique, Saclay, 91191 Gif sur Yvette, France\\}
\author{A. Tejeda}
\affiliation{Institut Jean Lamour,
CNRS - Nancy-Universit\'{e} - UPV-Metz, 54506 Vandouvre les Nancy, France}
\affiliation{Synchrotron SOLEIL, L'Orme des Merisiers, Saint-Aubin, 91192 Gif sur Yvette, France}
\author{A. Taleb-Ibrahimi}
\affiliation{UR1 CNRS/Synchrotron SOLEIL, Saint-Aubin, 91192 Gif sur Yvette, France}
\author{P. Le F\`{e}vre}
\affiliation{Synchrotron SOLEIL, L'Orme des Merisiers, Saint-Aubin, 91192 Gif sur Yvette, France}
\author{F. Bertran}
\affiliation{Synchrotron SOLEIL, L'Orme des Merisiers, Saint-Aubin, 91192 Gif sur Yvette, France}
\author{C. Berger}
\affiliation{Georgia Institute of Technology, Atlanta, Georgia
30332-0430, USA\\}\affiliation{CNRS/Institut N\'{e}el, BP166, 38042
Grenoble, France\\}
\author{W.A. de Heer}
\affiliation{Georgia Institute of Technology, Atlanta, Georgia
30332-0430, USA\\}
\author{A. Lanzara}
\affiliation{Dept. of Physics, Univ. of California, Berkeley, CA 94720, USA\\}
\affiliation{Advanced Light Source, Lawrence Berkeley National Laboratory, Berkeley, CA 94720, USA\\}
\author{E.H. Conrad}
\affiliation{Georgia Institute of Technology, Atlanta, Georgia
30332-0430, USA\\}

\begin{abstract}
Angle-resolved photoemission and X-ray diffraction experiments show that multilayer epitaxial graphene grown on the SiC$(000\bar{1})$ surface is a new form of carbon that is composed of effectively isolated graphene sheets. The unique rotational stacking of these films cause adjacent graphene layers to electronically decouple leading to a set of nearly independent linearly dispersing bands (Dirac cones) at the graphene $K$-point.  Each cone corresponds to an individual macro-scale graphene sheet in a multilayer stack where {\it AB}-stacked sheets can be considered as low density faults.
\end{abstract}
\vspace*{4ex}

\maketitle
\newpage

\begin{figure}
\includegraphics[width=5.5cm,clip]{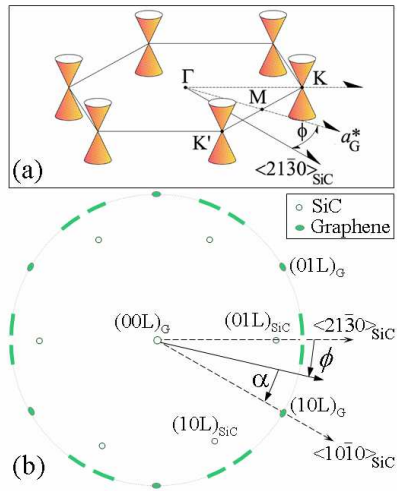}
\caption{(a) 2D Brillouin zone of graphene near $E_F$ showing the six Dirac cones at the $K$-points. The cones are shown rotated through an angle $\phi$ relative to the SiC $\langle 21\bar{3}0\rangle$ direction. (b) A schematic diffraction pattern of graphene grown on SiC$(000\bar{1})$.  The SiC diffraction pattern ($\circ$) and the graphene pattern ($\bullet$) from a $\phi = 30^\circ$ rotated film are shown.  Diffuse graphene arcs seen on C-face graphene are shown rotated $\phi\!\sim\!0^\circ$ from the $\langle21\bar{3}0\rangle$ direction.} \label{F:Brillouin_Z}
\end{figure}

The most fundamental property of an ideal graphene sheet is the linear dispersion of the $\pi$- and $\pi^*$-bands, $E(\Delta k)=\hbar v_F \Delta k$, where $v_F$ is the Fermi velocity and $\Delta k$ is the momentum relative to the $K$-points of the hexagonal reciprocal unit cell.\cite{graphene_BS_ref} The linear dispersion defines a cone with an apex at the Dirac point, $E_D$.\cite{graphene_BS_ref}  For undoped graphene, the Fermi Energy, $E_F$, equals $E_D$ so the Fermi surface consists of six points [see Fig.~\ref{F:Brillouin_Z}].
This unique dispersion is one of two fundamental properties that comprise the basis of an all-graphene electronics paradigm.\cite{Berger04}

Despite its importance for graphene physics, an unperturbed linear dispersion, especially near $E_D$, has not been directly observed. Exfoliated graphene flakes on $\text{SiO}_2$ have proven to be poor candidates for studying Dirac point physics because film disorder from impurities and mechanical deformation from substrate interactions cause huge position dependent charge fluctuations ($>\!10^{11}\text{cm}^{-2}$), implying that the Dirac cones are poorly defined for energies less than 0.3eV from $E_D$.\cite{Martin_NPYS_08} In fact, disorder induced broadening in exfoliated graphene makes the Dirac cone and the Dirac point unresolvable in Angle Resolved Photoemission (ARPES) experiments.\cite{Knox_PRB_08} The influence of the substrate has been somewhat reduced by suspending the films over microscopic holes,\cite{Du_NatureNano_08} however they remain susceptible to spontaneous rippling and strain.\cite{Meyer_SSCom_07}

In contrast epitaxial graphene (EG) grown directly on both the SiC(0001) Si-face and SiC$(000\bar{1})$ C-face has exceptional film quality. This coupled with its scalability to integrated circuits makes EG a serious material candidate for graphene electronics.\cite{Hass_JPhyCM_08,Emtsev_NatM_09} While disorder band broadening is not observed in EG, substrate interactions, as in exfoliated graphene, do play a role. Substrate interactions in Si-face graphene are known to distort the linear dispersion near $E_D$ in the first layer. They cause $\sim\!200$meV gap, up to $\sim\!500$meV electron doping, and enhanced electron-phonon coupling.\cite{Zhou_NatMat_07,Ohta_PRL_07,Bostwick_NatPH_07} Furthermore, the graphitic {\it AB}-stacking of Si-face graphene causes the band structure of these films to converge to graphite in thicker films.\cite{Ohta_PRL_07}

Of all forms of graphene (including single exfoliated sheets), only multilayer epitaxial graphene (MEG) grown on the C-face of SiC shows the essential signatures of an isolated graphene sheet; a Berry's phase of $\pi$, weak anti-localization, a square root dependence of the Landau level energies with applied magnetic field,\cite{Berger06,Sadowski06,Sadowski07a,Wu_PRL_07,de Heer_SSC_07,Orlita_PRL_08,Miller_Science_09} a zero landau level,\cite{Miller_Science_09} and as presented in this work, the unperturbed band structure of an isolated graphene sheet.

We present direct measurements of the linear band structure of MEG, explicitly demonstrating that it is \emph{not} graphite, but rather a new material consisting of essentially decoupled graphene layers.
We show that these films have long electron relaxation times and a remarkable absence of electron-phonon coupling or other distortions to the Dirac cone. These measurements demonstrate that a new periodic rotational stacking (\emph{not} the $60^\circ$ associated with graphite) is responsible for MEG's exquisite 2D properties.  These results support the theoretical explanation for preserving the linear bands.\cite{LopesdoSantos_PRL_07,Latil_PRB_07,Hass_PRL_08}

The substrates used in these studies were both n-doped $n\!=\!2\times\!10^{18}\text{cm}^{-2}$ 6H- and insulating 4H-SiC from Cree, Inc. Samples were prepared by $\text{H}_2$ treatments and subsequently grown in a closed RF induction furnace [see Ref.~[\onlinecite{Hass_JPhyCM_08}] for details]. The graphene films used in this study ranged from 11-12 layers as estimated by ellipsometry.\cite{Hass_JPhyCM_08} Samples were transported in air and thermally annealed at 800-1100\degC in UHV prior to measurement. It should be noted that furnace-grown samples have graphene domain sizes much larger than 20 microns, more than 100 times larger than graphene typically grown in UHV.\cite{Hiebel_PRB_08,Hass_JPhyCM_08}

ARPES measurements were made on different samples at both the Cassiop\'{e}e beamline at the SOLEIL synchrotron in Gif/Yvette and at the 12.0.1 beamline at the Advanced Light Source (ALS) at Lawrence Berkeley National Lab. The high resolution Cassiop\'{e}e beamline is equipped with a modified Peterson PGM monochromator with a resolution $E/\Delta E \simeq 70000$ at 100 eV and 25000 for lower energies. The detector is a $\pm 15^{\circ}$ acceptance Scienta R4000 detector with resolution $\Delta E\!<\!1$meV and $\Delta k\!\sim\!0.01\text{\AA}^{-1}$ at $\hbar \omega\!=\!30$ eV. Sample temperatures were varied from 6K to 300K.
The surface x-ray diffraction (SXRD) experiments were performed at the Advanced Photon Source, Argonne National Laboratory, on the 6IDB-$\mu$CAT UHV beam line with $\hbar \omega\!=\!16.2~$keV.

\begin{figure}[tp]
\includegraphics[width=8cm,clip]{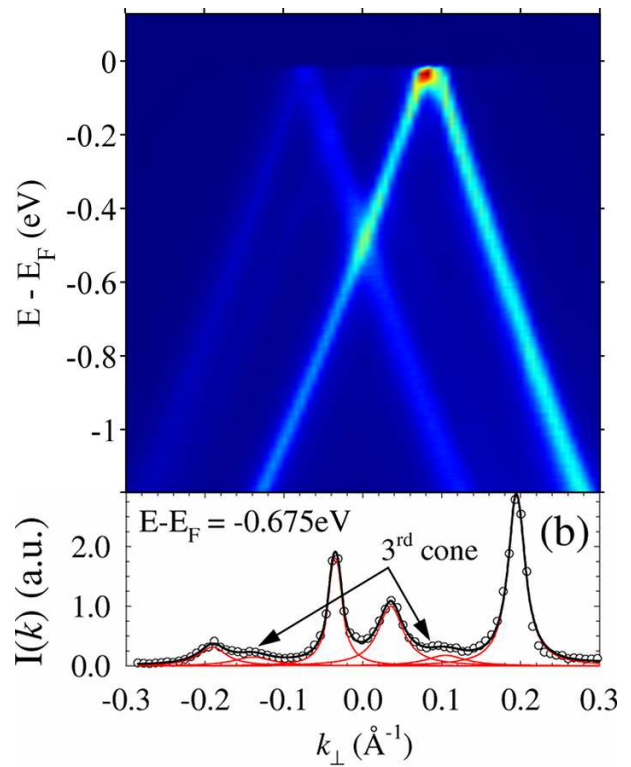}
\caption{(a) ARPES measured band structure of an 11-layer C-face graphene film grown on the 6H-SiC. The sample temperature is 6K. The scan is perpendicular to the SiC $\langle 10\bar{1}0\rangle_\text{SiC}$ direction at the $K$-point [See Fig.~\ref{F:Brillouin_Z}]. Three linear Dirac cones (one faint) are shown.  (b) An MDC  at $BE=E_F-0.675$eV shows all three cones.  Heavy solid line is a fit to the sum of six Lorentzians (thin solid lines).} \label{F:K_PointBS}
\end{figure}
The primary result of this work is shown in Fig.~\ref{F:K_PointBS}(a), where we display the band structure of an 11-layer graphene film grown on the C-face of 6H-SiC.  Data is taken near the $K$-point ($k_{\Gamma K}\!=\!1.704\text{\AA}^{-1}$, $k_z\!\sim\!0.02c^*$, where $c^*\!=\!2\pi/6.674\text{\AA}\!=\! 0.941\text{\AA}^{-1}$) and not at the $H$-point of graphite ($k_z\!\sim\!0.5c^*$). The figure shows two bright and one faint intersecting Dirac cones; the faint cone is more easily visible in the Momentum Dispersion Curve (MDC) in Fig.~\ref{F:K_PointBS}(b). The Dirac cones in Fig.~\ref{F:K_PointBS}(a) are the first measured unperturbed $\pi$-bands expected from an isolated graphene sheet. Band maps on different samples and different parts of the sample show similar results: multiple rotated linearly dispersing Dirac cones. Within the experimental uncertainty ($\sim\!20$meV), there is no evidence of a band gap. Because ARPES is sensitive to 3-4 surface layers at this energy, there is no influence on the bands from the graphene-SiC substrate. For this reason, the difference between the Dirac point and $E_F$ is $0.0 \pm 20$meV. This puts an upper limit on the doping level to be $<10^{10}\text{cm}^{-2}$, consistent with IR measurements from similar films ($5\!\times\!10^9\text{cm}^{-2}$).\cite{Orlita_PRL_08}

Two points must be stressed. First, these films are \emph{not} graphitic. The {\it AB}-stacking of graphite would show parabolic bands~\cite{Zhou_Nat_06} or the splitting seen in bilayer or multilayer graphene films grown on the Si-face of SiC.\cite{Zhou_NatMat_07,Ohta_PRL_07}  In fact, {\it AB} planes are so few in C-face MEG films that they can be viewed as stacking faults in these films. The second point that must be kept in mind is that furnace-grown and UHV-grown graphene are very different, both structurally and electronically. In addition to a two order of magnitude reduction in graphene domain size, ARPES measurements on UHV-grown C-face graphene show a large electron doping of $E_D\!-\!E_F\!=\!0.2$eV with poorly developed $\pi$- and $\sigma$-bands.\cite{Emtsev_PRB_08} The doping level difference is likely due to charge coupling between the SiC and the thinner UHV films, while the broad $\pi$-bands are due to film disorder. The remarkable result of multiple linear bands characteristic of rotated but isolated single graphene sheets confirms predictions that the unique stacking of multilayer graphene grown on the C-face of SiC preserves the symmetry of isolated graphene.\cite{LopesdoSantos_PRL_07,Latil_PRB_07,Hass_PRL_08} To demonstrate this we first point out a few structural details of C-face films.

\begin{figure}[ht]
\includegraphics[width=7cm,clip]{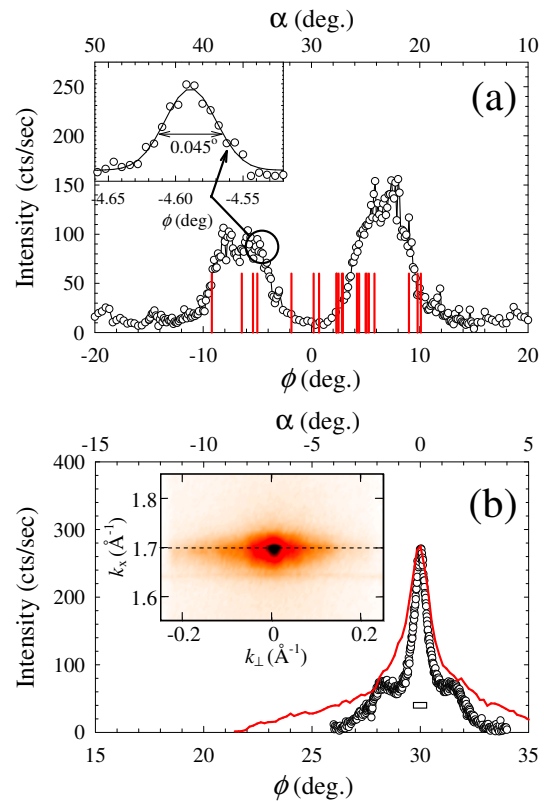}
\caption{(a) SXRD angular distribution of the diffuse arcs in Fig.~\ref{F:Brillouin_Z}(b).  Insert in (a) shows  a magnified view of a single rotation angle. Solid lines mark the angular position $\alpha$ [upper scale] of measured ARPES Dirac cones relative to the $\langle 10\bar{1}0\rangle$ direction. (b) ($\bigtriangleup$) SXRD angular distribution around $\phi=30^\circ$. Insert in (b) is constant energy cut at $E_D(k_x=k_{\Gamma K})$ showing the distribution of cones perpendicular to $\Gamma K$ (i.e. in $k_\perp$).  Solid line in (b) is a plot of the angular distribution of Dirac cones versus $\alpha$ (upper scale). Rectangle in (b) shows the ARPES angular resolution.} \label{F:Diffraction}
\end{figure}

We have plotted SXRD azimuthal scans near $\phi\!=\!0^\circ$ and $30^\circ$ in Fig.~\ref{F:Diffraction}. Note that, while the exact distribution of graphene rotation angles is sample dependent, the probability of rotation angles near $\phi\!=\!30^\circ$ is nearly equal to the probability of angles near $0^\circ$, regardless of sample or film thickness (i.e., the area under the x-ray curves are nearly equal). This, along with SXRD reflectivity measurements, implies that approximately every other sheet is rotated $\sim\!30^\circ$ instead of the graphitic $\sim\!60^\circ$,~\cite{Hass_PRL_08,Hass_JPhyCM_08} and not the ``occasional'' small angles rotations proposed by STM measurements.\cite{Biedermann_PRB_09}  In other words, {\it AB} pairs should be considered to be faults in the stacking order. The distribution of rotation angles around $\phi\!=\!0^\circ$ and $30^\circ$ is determined by an entropy term that selects from a number of SiC-graphene commensurate angles with small energy differences.\cite{Hass_JPhyCM_08} There are more commensurate angles per radian of arc at $\phi\!=\!0^\circ$, which explains the observed broader distribution around $0^\circ$ in Fig.~\ref{F:Diffraction}(a).\cite{Hass_JPhyCM_08} Also note that the angular width of each discrete rotation is very narrow; a detailed scan of one such angle is shown in the insert of Fig.~\ref{F:Diffraction}(a). Its width is $0.045^\circ$, corresponding to an x-ray rotational coherence distance of $\sim\!1\mu$m.

To show the correlation between graphene rotation angle $\phi$ and the $\Gamma K$ rotation direction $\alpha$, note that the $\Gamma K$ direction in ARPES is rotated $30^\circ$ from the graphene reciprocal space direction, $a^*_G$ [see Fig.~\ref{F:Brillouin_Z}(a)]. This means that the $\Gamma K$ direction for a graphene sheet rotated $\phi$ from the $\langle\!21\bar{3}0\!\rangle$ direction is at an angle $\alpha\!=\!\phi\!-\!30^\circ$ relative to the SiC $\langle\!10\bar{1}0\!\rangle$ [see Fig.~\ref{F:Brillouin_Z}]. For example, graphene rotated $\phi\!=\!30^\circ$ relative to the $\langle\!21\bar{3}0\!\rangle$ direction of SiC has the $\Gamma K$ direction along the $\langle\!21\bar{3}0\!\rangle$ direction. We have marked the discrete rotation angles of the ARPES Dirac cones (near $\alpha\!=\!30^\circ$) against the angular distribution measured by SXRD in Fig.~\ref{F:Diffraction}(a) [$\alpha\!=\!30^\circ+\text{tan}^{-1}(k_\perp /k_{\Gamma K})$, where $k_\perp$ is taken from ARPES scans like the one shown in Fig.~\ref{F:K_PointBS}]. It is clear that the rotated cones correlate well with the data with many more rotations between $2^\circ$ and $10^\circ$. Note that the SXRD beam size is $\sim\!3$mm while the ARPES beam size is $\sim\!40\mu$m; this is why ARPES data shows a small number of discrete rotated cones and SXRD shows a more continuous distribution averaged over a large beam footprint. In the $\alpha\!=\!0^\circ$ azimuth discrete cones are not always resolved because of the narrow rotational distribution as seen in the inset in Fig.~\ref{F:Diffraction}(b).  Note that angular scale in Fig.~\ref{F:Diffraction}(b) is expanded by a factor of 2 compared to (a). The reason discrete cones are not observed is a combination of the narrow distribution of commensurate rotations at $\phi\!=\!30^\circ$ and the wide angular acceptance of ARPES ($\sim\!0.34^\circ$). Nonetheless, the ARPES distribution of cones again coincide with the measured SXRD angular distribution [Fig.~\ref{F:Diffraction}(b)].

\begin{figure}[ht]
\includegraphics[width=7.5cm,clip]{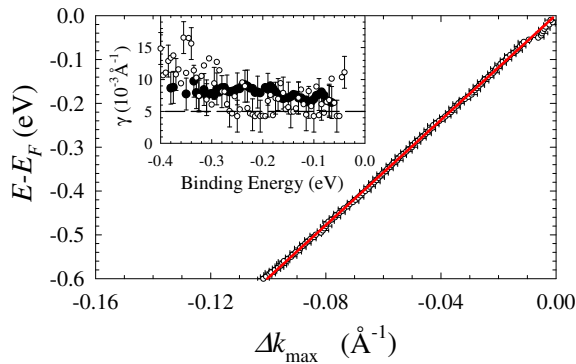}
\caption{$E(\Delta k_\text{max})\!-\!E_F$ versus $\Delta k\!=\!k_D\!-\!k$. $k_D$ is the $K$-point position and $k$ is the Lorentzian center from fits to ARPES MDCs.  $\Delta k_\text{max}$ is measured perpendicular to the $\Gamma K$ direction near the $K$-point. Solid line is a linear fit. Insert is a plot of the MDC HWHM, $\gamma$, as a function of binding energy at 6K ($\bullet$) and 300K ($\circ$).  Dashed line in the ARPES resolution.} \label{F:Width}
\end{figure}

The high energy resolution dispersion curves allow us to measure two important effects.  First, the bands are linear. This is demonstrated more clearly in Fig.~\ref{F:Width} where we plot the position of one branch of a Dirac cone (determined by fitting the ARPES MDCs to Lorentzian peaks).  Within the error bars of the experiment, there are no significant deviations from linearity, consistent with weak electron-phonon coupling at very low carrier densities.\cite{Calandra_PRB_07} The average Fermi velocity, derived from the slope of $E(\Delta k)$, was found to be $\langle v_F\rangle\!=\!1.0\pm 0.05\times 10^6$m/sec for energies down to $\sim 0.5$eV below $E_D$.  This value is larger than $v_F$ for bulk graphite ($v_F\simeq 0.86\times10^6$m/sec)~\cite{Bychkov_PRB_08} but within error bars of values obtained from both IR measurements ($1.02\pm 0.01\times 10^6$m/s)\cite{Orlita_PRL_08} and scanning tunneling spectroscopy ($1.07\pm 0.01\times 10^6$m/s).\cite{Miller_Science_09}

The second point to note is the narrow Lorentzian half width at half maximum ($\gamma$) of an MDC [insert in Fig.~\ref{F:Width}]. $\gamma$ is inversely proportional to the carrier scattering time $\tau =1/(2\gamma v_F)$.\cite{Calandra_PRB_07} Because $\gamma$ is within error bars of the instrument resolution, we are only able to place a lower bound of $\tau\!>\!20$fs.  This is consistent with $\tau$ from IR measurements (100 - 300fs).\cite{Orlita_PRL_08} Also note that there is no measurable change in $\tau$ between 6K and 300K.



ARPES measurements show that the band structure of MEG graphene grown on the C-face of SiC consists of multiple undistorted, linearly dispersing graphene bands originating from individual rotated layers in the multilayer film. The observed Dirac cones definitively demonstrate that that the graphene sheets in the MEG films can be considered as electronically ideal isolated graphene sheets. The origin of this unique behavior is a result of MEG's unique stacking order. All that is required to preserve graphene's linear dispersion in a multilayer stack is to break the {\it AB}-stacking symmetry of graphite.  This is realized by introducing a relative rotation angle between two adjacent sheets that is not $60^\circ$ (i.e. graphite stacking).\cite{LopesdoSantos_PRL_07,Latil_PRB_07,Hass_PRL_08} As C-face graphene films grow, the substrate apparently forces relative rotation of $\sim\!30\!\pm\!7^\circ$ making graphitic {\it AB}-stacked pairs infrequent faults in the film. The significance of this result is that uniform single or double-layer graphene films are not necessarily a requirement for graphene electronics, since even multilayer films have the required electronic properties.

This research was supported by the W.M. Keck Foundation, the Partner University Fund from the Embassy of France, and the NSF under Grant No. DMR-0820382.  The Advanced Photon Source is supported by the DOE Office of BES, contract W-31-109-Eng-38.  The $\mu$-CAT beam line is supported by the US DOE through Ames Lab under Contract No.W-7405-Eng-82.


\begin{thebibliography}{99}


\bibitem{graphene_BS_ref}P.R. Wallace, {\it Phys. Rev.} {\bf 71}, 622 (1947); J.W. McClure, {\it Phys. Rev.} {\bf 108}, 612 (1957).
\bibitem{Berger04}C. Berger, et al.,
{\it J. Phys. Chem.} {\bf B 108}, 19912 (2004).
\bibitem{Martin_NPYS_08} J. Martin, N. Akerman, G. Ulbricht, T. Lohmann, J. H. Smet, K.
von Klitzing, and A. Yacoby, {\it Nat. Phys.} {\bf 4}, 144 (2008).
\bibitem{Knox_PRB_08}K.R. Knox, S. Wang, A. Morgante, D. Cvetko, A. Locatelli, T.O. Mentes, M.A. Ni\~{n}o, P. Kim, and R.M. Osgood Jr., {\it Phys. Rev.} {\bf B 78}, 201408(R)(2008).
\bibitem{Du_NatureNano_08}X. Du, I. Skachko, A. Barker, E.Y. Andrei, {\it Nature Nano.} {\bf 3}, 491 - 495 (2008).
\bibitem{Meyer_SSCom_07}J.C Meyer, A.K. Geim, M.I. Katsnelson, K.S. Novoselov, D. Obergfell, S. Roth, C. Girit and A. Zettl, {\it Solid State Com.} {\bf 143} 101 (2007).
\bibitem{Hass_JPhyCM_08}J. Hass, W.A. de Heer and E.H. Conrad, {\it J. Phys.: Condens. Matt.} {\bf 20}, 3232002 (2008).
\bibitem{Emtsev_NatM_09}K.V. Emtsev, et al.,
{\it Nature Mat.} {\bf 8}, 203 (2009).

\bibitem{Zhou_NatMat_07}S.Y. Zhou, G.-H. Gweon, A.V. Fedorov, P.N. First, W.A. De Heer, D.-H. Lee, F. Guinea, A.H. Castro Neto AND A. Lanzara, {\it Nature Mat.} {\bf 6} 770 (2007).
\bibitem{Ohta_PRL_07}T. Ohta, A. Bostwick, J.L. McChesney, T. Seyller, K. Horn and E. Rotenberg, {\it Phys. Rev. Lett.} {\bf 98}, 206802 (2007).
\bibitem{Bostwick_NatPH_07} A. Bostwick, T. Ohta, T. Seyller, K. Horn and E. Rotenberg, {\it Nature Phys.} {\bf 3}, 36 (2007).
\bibitem{Orlita_PRL_08}M. Orlita, C. Faugeras, P. Plochocka, P. Neugebauer, G. Martinez, D.K. Maude, A.-L. Barra, M. Sprinkle, C. Berger, W.A. de Heer, and M. Potemski, {\it Phys. Rev. Lett.} {\bf 101}, 267601 (2008).
\bibitem{Berger06}C. Berger, Z. Song, X. Li, X. Wu, N. Brown, C. Naud, D. Mayou, T. Li, J. Hass, A.N. Marchenkov, E.H. Conrad, P.N. First and W.A. de Heer, {\it Science} {\bf 312}, 1191 (2006).
\bibitem{Sadowski06} M.L. Sadowski and G. Martinez and M. Potemski and C. Berger and W.A. de Heer, {\it Phys. Rev. Lett.} {\bf 97}, 266405 (2006).
\bibitem{Sadowski07a}M.L. Sadowski, G. Martinez, M. Potemski, C. Berger and W.A. de Heer, {\it Solid State Commun.} {\bf 143}, 123 (2007).
\bibitem{Wu_PRL_07} X. Wu and X. Li and Z. Song and C. Berger and W.A. de Heer, {\it Phys. Rev. Lett.} {\bf 98}, 136801 (2007).
\bibitem{de Heer_SSC_07}W.A. de Heer, et al.,
{\it Sol. State Comm.} {\bf 143}, 92 (2007).
\bibitem{Miller_Science_09} D.L. Miller, K.D. Kubista, G.M. Rutter, M. Ruan, W.A. de Heer, P.N. First and J.A. Stroscio., {\it Science}, {\bf 324} 924 (2009).
\bibitem{LopesdoSantos_PRL_07}J.M.B. Lopes dos Santos, N.M.R. Peres and A.H. Castro Neto, {\it Phys. Rev. Lett.} {\bf 99}, 256802 (2007).
\bibitem{Latil_PRB_07}S. Latil, V. Meunier and L. Henrard, {\it Phys. Rev.} {\bf B 76}, 201402(R) (2007).
\bibitem{Hass_PRL_08}J. Hass, F. Varchon, J.E. Millan-Otoya, M. Sprinkle, N. Sharma, W. A. de Heer, C. Berger, P. N. First, L. Magaud and E.H. Conrad, {\it Phys. Rev. Lett.} {\bf 100}, 125504 (2008).
\bibitem{Hiebel_PRB_08}F. Hiebel, P. Mallet, F. Varchon, L. Magaud, and J-Y. Veuillen, {\it Phys. Rev.} {\bf B 78}, 153412 (2008).
\bibitem{Zhou_Nat_06}S.Y. Zhou, G.-H. Gweon, J. Graf, A.V. Fedorov, C.D. Spataru, R.D. Diehl, Y. Kopelevich, D.-H. Lee, S.G. Louie and A. Lanzara, {\it Nature Physics} {\bf 2}, 595 (2006).
\bibitem{Emtsev_PRB_08}K. V. Emtsev, F. Speck, Th. Seyller, and L. Ley, and J. D. Riley, {\it Phys. Rev.} {\bf B 77}, 155303 (2008).
\bibitem{Biedermann_PRB_09}L.B. Biedermann, M.L. Bolen, M.A. Capano, D. Zemlyanov, and R.G. Reifenberger, {\it Phys. Rev.} {\bf B 79}, 125411 (2009).
\bibitem{Calandra_PRB_07}M. Calandra and F. Mauri, {\it Phys. Rev.} {\bf B 76}, 205411 (2007)
\bibitem{Bychkov_PRB_08}Yu. A. Bychkov and G. Martinez, {\it Phys. Rev.} {\bf B 77}, 125417 (2008).










\end{thebibliography}
\end{document}